\documentclass[lettersize,journal]{IEEEtran}
\usepackage{amsmath,amsfonts}
\usepackage{algorithmic}
\usepackage{algorithm}
\usepackage{array}
\usepackage[caption=false,font=normalsize,labelfont=sf,textfont=sf]{subfig}
\usepackage{textcomp}
\usepackage{stfloats}
\usepackage{url}
\usepackage{verbatim}
\usepackage{graphicx}
\usepackage{cite}
\usepackage{tikz}
\usetikzlibrary{shapes.geometric, arrows.meta, positioning,calc}
\usepackage{amsfonts}
\usepackage{graphicx}
\usepackage{subcaption}
\usepackage[]{glossaries}
\usepackage{etoolbox}
\usepackage{subcaption}

\usepackage{pgfplots}
\pgfplotsset{compat=newest}

\usepackage{background}

\SetBgAngle{90}
\SetBgPosition{current page.east}
\SetBgVshift{6mm}
\SetBgHshift{0mm}
\SetBgColor{black}
\SetBgOpacity{1}
\SetBgScale{1.5}

\SetBgContents{%
	\parbox{15cm}{\scriptsize\raggedright
		This work has been submitted to the IEEE Transactions on Neural Networks and Learning Systems for possible publication.\\
		Copyright may be transferred without notice, after which this version may no longer be accessible.%
	}%
}

\newcommand{\shortminus}{\scalebox{0.8}[1.0]{$-$}}
\newcommand{\shortplus}{\scalebox{0.8}[0.8]{$+$}}
\newglossary[slg]{symbols}{sls}{slo}{List of Symbols}
\newglossaryentry{L_RNN}{
	name={\text{$L_{\mathrm{RNN}}$}},
	description={Analytical upper bound for Lipschitz constant},
	sort=LRNN,
	type=symbols
}

\newglossaryentry{0}{
	name={\text{$\mathbf{0}$}},
	description={Matrix with all entries being zero},
	sort=zero,
	type=symbols
}
\newglossaryentry{L_act}{
	name={\text{$L_{\mathrm{act}}$}},
	description={Lipischtz constant estimated via active search},
	sort=LRNN,
	type=symbols
}
\newglossaryentry{L_rand}{
	name={\text{$L_{\mathrm{rand}}$}},
	description={Lipischtz constant estimated via random sampling},
	sort=LRNN,
	type=symbols
}
\newglossaryentry{n_in}{
	name={\text{$n_{\mathrm{in}}$}},
	description={Size of the input vector $u_t$},
	sort=-nin,
	type=symbols
}
\newglossaryentry{n_out}{
	name={\text{$n_{\mathrm{out}}$}},
	description={Size of the output vector $y_t$},
	sort=-nout,
	type=symbols
}
\newglossaryentry{n_h}{
	name={\text{$n_\mathrm{h}$}},
	description={Size of the hidden layer(s)},
	sort=-nu,
	type=symbols
}
\newglossaryentry{x_t}{
	name={\text{$\mathbf{x}_t$}},
	description={Input vector at timestep $t$},
	sort=ut,
	type=symbols
}
\newglossaryentry{u}{
	name={\text{$\tilde{\mathbf{u}}$}},
	description={Concatenation of the input vectors over $T$ time steps},
	sort=uz,
	type=symbols
}
\newglossaryentry{x_i}{
	name={\text{$\mathbf{x}_i$}},
	description={Activation of a feed forward neural network at layer $i$},
	sort=xi,
	type=symbols
}
\newglossaryentry{x}{
	name={\text{$\tilde{\mathbf{x}}$}},
	description={Concatenation of the activations of a feed forward neural network},
	sort=xx,
	type=symbols
}
\newglossaryentry{y_t}{
	name={\text{$\mathbf{y}_t$}},
	description={Output vector at timestep t},
	sort=yt,
	type=symbols
}
\newglossaryentry{y}{
	name={\text{$\tilde{\mathbf{y}}$}},
	description={Concatenation of the output vectors over $T$ time steps},
	sort=yz,
	type=symbols
}
\newglossaryentry{h_t}{
	name={\text{$\mathbf{h}_t$}},
	description={Hidden state vector of a recurrent neural network at t},
	sort=ht,
	type=symbols
}

\newglossaryentry{h}{
	name={\text{$\tilde{\mathbf{h}}$}},
	description={Concatenation of the hidden state vectors of a recurrent neural network over $T$ time steps},
	sort=hz,
	type=symbols
}
\newglossaryentry{W_i}{
	name={\text{$\mathbf{W}_i$}},
	description={Weight matrix of the  $i$-th feed-forward layer},
	sort=Wai,
	type=symbols
}
\newglossaryentry{W_h}{
	name={\text{$\mathbf{W}_{\mathrm{h}}$}},
	description={Recurrent weight matrix mapping the hidden state vector from the previous time step to the current hidden state vector.},
	sort=Whh,
	type=symbols
}
\newglossaryentry{W_x}{
	name={\text{$\mathbf{W}_{\mathrm{x}}$}},
	description={Weight matrix mapping input vector to the hidden state vector in the RNN},
	sort=Wih,
	type=symbols
}
\newglossaryentry{W_out}{
	name={\text{$\mathbf{W}_{\mathrm{out}}$}},
	description={Weight matrix of the output layer},
	sort=Waout,
	type=symbols
}
\newglossaryentry{b_i}{
	name={\text{$\mathbf{b}_i$}},
	description={Bias vector of the  $i$-th layer},
	sort=bi,
	type=symbols
}
\newglossaryentry{b_out}{
	name={\text{$\mathbf{b}_{\mathrm{out}}$}},
	description={Bias vector of the output layer},
	sort=bout,
	type=symbols
}
\newglossaryentry{z}{
	name={\text{$\tilde{\mathbf{z}}$}},
	description={Joined state vector of a RNN. Consisting of the inputs $\tilde{\mathbf{x}}$ and hidden states $\tilde{\mathbf{h}}$},
	sort=z,
	type=symbols
}
\newglossaryentry{T}{
	name={\text{$\mathbf{T}$}},
	description={Diagonal decision matrix for constraint},
	sort=T,
	type=symbols
}
\begin{document}

\title{Lipschitz-Based Robustness Certification for Recurrent Neural Networks via Convex Relaxation}

\author{Paul Hamelbeck and Johannes Schiffer}


\markboth{}%
{Shell \MakeLowercase{\textit{et al.}}: A Sample Article Using IEEEtran.cls for IEEE Journals}


\maketitle

\begin{abstract}
	Robustness certification against bounded input noise or adversarial perturbations is increasingly important for deployment recurrent neural networks (RNNs) in safety‑critical control applications. To address this challenge, we present RNN-SDP, a relaxation based method that models the RNN's layer interactions as a convex problem and computes a certified upper bound on the Lipschitz constant via semidefinite programming (SDP). We also explore an extension that incorporates known input constraints to further tighten the resulting Lipschitz bounds. RNN-SDP is evaluated on a synthetic multi-tank system, with upper bounds compared to empirical estimates. While incorporating input constraints yields only modest improvements, the general method produces reasonably tight and certifiable bounds, even as sequence length increases. The results also underscore the often underestimated impact of initialization errors—an important consideration for applications where models are frequently re-initialized, such as model predictive control (MPC).
\end{abstract}


\section{Introduction}

\IEEEPARstart{N}{eural} networks (NNs) have demonstrated remarkable success across a wide range of applications, including computer vision~\cite{He.2016}, natural language processing~\cite{Manning.2014}, and control systems~\cite{Narendra.1996}, making them indispensable tools in modern machine learning and artificial intelligence. However, their highly non-linear and high-dimensional structure often renders them opaque, leading to their characterization as black-box models: capable of impressive performance, but offering limited interpretability or formal guarantees regarding their behavior~\cite{Liu.2021}. This opacity becomes particularly problematic in safety-critical domains, such as autonomous driving, medical diagnostics, or large-scale control systems, where even small input perturbations can result in unpredictable or unsafe outcomes~\cite{Szegedy.21122013}. In such settings, formal verification and robustness analysis are essential to ensure reliability, stability, and trust in NN-based decision systems\cite{Liu.2021}.

In response to these challenges, substantial efforts have been made to reduce noise sensitivity, while maintaining the predictive capabilities NNs as well as to develop tools for rigorously assessing their robustness. Techniques such as adversarial training~\cite{Zheng.2016} and regularization~\cite{Wu.2021} aim to reduce a network’s sensitivity to input perturbations and thereby increas robustness against noise and adversarial attacks. While these robustness-enhancing techniques play a critical role during training, they must be complemented by verification tools that can formally assess the model’s behavior across a range of inputs. Such verification techniques typically fall into one of three categories: \emph{a) reachability-based}, \emph{b) optimization-based}, or \emph{c) relaxation-based} verification methods, each offering a different balance of precision, scalability, and generality.

\paragraph{Reachability-based} Reachability-based verification methods aim to compute the full set of possible outputs $\mathcal{Y}$ that a NN can produce for all inputs $\mathcal{X}$ within a given region. This is typically achieved through layer-by-layer propagation of the input set, which can be computationally expensive due to the exponential growth in the number of activation regions as the network depth increases. Methods that do not introduce simplifications yield exact reachable sets without over-approximation, but their scalability is limited to small networks~\cite{Xiang.2017}. To address this, many approaches incorporate set-based relaxations or over-approximations, which significantly reduce computational complexity and make the analysis tractable for larger models~\cite{Gehr.2018}. Additionally, reachability-based methods as a whole are generally limited to networks with ReLU or other piecewise-linear activation functions, as these preserve the geometric structure necessary for sound and efficient set propagation~\cite{Liu.2021}. 
\IEEEpubidadjcol
\paragraph{Optimization-based}Optimization-based verification methods attempt to either find a counterexample, that is within the set $\mathcal{U}$ that produces an output outside the safe region $\mathcal{Y}$~\cite{Lomuscio.2017} or identify the worst-case input that causes the most undesirable behavior, as defined by a given cost function~\cite{Tjeng.20112017}. These approaches commonly encode NNs as a set of mixed-integer linear constraints, enabling formal optimization. While such formulations can be solved using complete solvers like branch-and-bound, which can either find a violation or prove that none exists, they are typically limited to networks with piecewise-linear activation functions, such as ReLU or max-pooling~\cite{Tjeng.20112017}. Another commonly used solver, the satisfiability modulo theories (SMT) solver, can, in principle, also handle arbitrary non-linear activation functions (e.g., sigmoid or tanh). However in practice applications to networks with such activation functions remain limited as computational costs drastically increase and the completness of the approach is lost~\cite{Guidotti.2023}.

\paragraph{Relaxation-based}Relaxation-based verification methods approximate the constraints imposed by the NN with looser, more tractable formulations, typically to enable scalable analysis~\cite{Liu.2021}. Instead of precisely encoding every activation or network branch, these methods relax non-linear or combinatorial behaviors into convex or piecewise-linear forms. The result is typically a bound on the network's output or sensitivity, rather than a definitive yes/no certificate. The central trade-off is between tightness of the bound and computational efficiency: coarser relaxations scale well but offer weaker guarantees, while tighter relaxations are more informative but expensive to compute. Prominent examples include: CROWN~\cite{HuanZhang.2018}, which propagates linear bounds layer-by-layer; ConvDual~\cite{EricWong.2018}, which uses dual feasible solutions to certify output ranges; and LipSDP~\cite{fazlyab2023efficientaccurateestimationlipschitz}, which leverages a convex optimization framework, that yields certified upper bounds on the global Lipschitz constant on a  feed-forwards neural network (FFNN) by approximating non-linear activation functions with quadratic constraints. 

The LipSDP approach is particularly appealing from a control-theoretic perspective, as the Lipschitz constant, which bounds the worst-case change in output relative to input variation, is a widely used robustness metric in many control frameworks, such as observer design~\cite{Rajamani.1998}, and model predictive control (MPC)~\cite{Yu.2013}. However, recent advances in system identification and control have seen a shift from FFNNs to recurrent neural networks (RNNs), due to their superior capacity for modeling temporal dynamics~\cite{BONASSI2022_review,Wong.2018,Mohajerin.2019}. RNNs achieve this by incorporating an internal feedback loop, which enables the retention of historical information and facilitates sequential prediction. In contrast, FFNNs generate outputs based solely on the current input. While this internal recurrence enhances the representational capacity of RNNs for dynamic systems, it also introduces structural complexity that renders many existing verification techniques inapplicable. Bonassi et al.~\cite{BONASSI2022_review} specifically highlight the absence of a method akin to LipSDP, in the context of RNNs, noting its importance for control applications that demand certified robustness.

\paragraph*{Our Contributions} We propose the RNN-SDP framework for tightly bounding the Lipschitz constant of RNNs, enabling robustness and stability analysis. Inspired by LipSDP~\cite{fazlyab2023efficientaccurateestimationlipschitz}, we propose an SDP-based certifiaction techique tailored to RNNs, producing finite‑horizon input–output Lipschitz bounds that account for initialization errors
 The latter is particularly critical in applications with frequent re-initialization, which prevents the system from reaching a steady state, making initialization effects persistently relevant. Additionally, we present a framework to utilize known input constraints, which are commonly present in practical NN applications~\cite{Fergus2022}, to further tighten the estimated bound of the Lipschitz constant.

\paragraph*{Our Approach} By restricting the sequence length for predictions to an arbitrary finite length of $N$,  we can transform the RNN into an equivalent FFNN via unrolling~\cite{Miller.25052018}. For this unrolled structure, we exploit the slope-restricted nature of non-linear activation functions to derive an upper bound on the Lipschitz constant through semidefinite programming (SDP). To incorporate input constraints, we refine global slope bounds into local slope restrictions for each neuron, informed by the constrained input domain and the inter-layer interactions.

\paragraph*{Our Results} We evaluate our method on RNNs trained to predict tank levels in a multi-tank system. We compare our Lipschitz bound against estimates obtained via random sampling and active exploration. Our results highlight the potential risks of relying solely on statistical estimation and demonstrate the tightness of our approach across different sequence lengths. For short sequences, where the combined impact of input and initialization disturbances is most pronounced, our method provides bounds roughly 1\% above the empirical lower bound estimates, offering strong worst-case guarantees. For longer sequences, the bound loosens to about 30\%, reflecting the inherent challenge of tight bounding over extended horizons while still maintaining provable robustness.

\paragraph*{Related Work} Lipschitz constant estimation for RNNs is less mature than for feed-forward architectures, but several approaches have emerged. Revay et al.~\cite{revay2020convexparameterizationrobustrecurrent} introduced a convex parameterization framework for RNNs that provides robustness and stability guarantees during training. This was later extended into the recurrent equilibrium network (REN) architecture~\cite{revay2024}, which models RNNs as feedback interconnections of linear systems and monotonic non-linearities. Their method uses incremental quadratic constraints to certify global exponential stability and bounded incremental $L_2$-gain—effectively a Lipschitz bound on the sequence-to-sequence mapping. However, their framework does not account for the impact of initialization. While effective for general system identification, this method tends to be less suitable frequently re-initialized models, where accurate treatment of initialization errors is critical. A similar limitation can be seen in the method of Guo et al.~\cite{Guo_10.1145/3583788.3583795}. They provide tight and validated Lipschitz bounds by leveraging interval arithmetic and Clarke-gradient enclosures. This leads to rigorous certification, especially valuable in low-dimensional short sequence data. Their approach assumes a fixed initial-state and therefore does also not account, for initialization impacts. Our proposed approach (RNN-SDP) accounts for these effects and therefore provides a valuable addition to the currently available tool-set for robustness certification, especially for frequently re-initialized applications, such as MPC~\cite{Giuli10885933}.

\section{Preliminaries}
In this section, we introduce the notation and provide the definitions and properties of RNNs relevant to this work: RNN layer formulation, stability considerations, and the unrolled representation.

\subsection{Notation}
Vectors are denoted by lowercase bold letters (e.g., $\mathbf{x}$), and matrices by uppercase bold letters (e.g., $\mathbf{W}$). Stacked vectors—i.e., vectors formed by vertically concatenating multiple time- or layer-indexed vectors—are denoted by a tilde (e.g., $\tilde{\mathbf{x}}$). The $n$-dimensional identity matrix is denoted by $\mathbf{I}_n$, while the zero matrix, i.e., a matrix of appropriate dimension with all entries equal to zero, is denoted by \gls{0}. The notation $\operatorname{diag}(a_1, \dots, a_n)$ refers to a diagonal matrix with diagonal entries $a_1$ through $a_n$, starting from the upper-left corner. The $p$-norm of a vector, for $p \geq 1$, is denoted by $||\cdot||_p : \mathbb{R}^n \to \mathbb{R}_0^+$. The notation $[a, b]$ denotes a closed interval for scalar values or component wise bounds when used with vectors. 

\subsection{RNN Layer Definition}
\label{subsec:RNNDefinition}
We briefly recall the structure of a standard RNN in order to establish our employed notation based on \cite{Sutskever.2011}. 
An RNN maintains an internal \emph{hidden state} $\mathbf{h}_t \in \mathbb{R}^n$ that serves as a memory of past inputs and influences future predictions. 
At each discrete time step $t \in \mathbb{Z}_{\ge 0}$, the hidden state is updated according to
\begin{equation}
	\mathbf{h}_{t} = \phi\!\left(\mathbf{W}_h \mathbf{h}_{t-1} + \mathbf{W}_x \mathbf{x}_{t} + \mathbf{b}\right),
	\label{eq:RNNupdate}
\end{equation}
where $\mathbf{W}_h \in \mathbb{R}^{n\times n}$ applies a linear transformation to the previous hidden state, $\mathbf{W}_x \in \mathbb{R}^{n\times m}$ transforms the input $\mathbf{x}_{t} \in \mathbb{R}^m$, $\mathbf{b} \in \mathbb{R}^n$ is a bias vector, and $\phi(\cdot)$ denotes a element wise non-linear activation function, such as the hyperbolic tangent $\tanh(\cdot)$ used in classical RNNs.

\begin{figure*}[!t]
	\centering
	\begin{tikzpicture}[
		actFunc/.style={rectangle, draw=black, minimum width=2.8cm, minimum height=1.2cm, align=center},
		>=Latex
		]
		
		\def\arrowLength{1.8cm};
		\def\textOffset{0.35cm};
		
		\node[actFunc] (L1) at (3,0) {$\phi(\gls{W_x} \mathbf{x}_{t} + \gls{W_h} \mathbf{h}_{t-1} + \mathbf{b})$};
		\node[actFunc] (Lout) at ($(L1.east)+(3.5,0)$) {$\gls{W_out} h_{t} + \gls{b_out}$};

		\draw[->] (L1.south) |- ($(L1.south)+(-2.5,-1)$) -- ($(L1.north)+(-2.5,1)$)-|($(L1.north)$);
		\draw [draw=white,fill=white] ($(L1)+(-2.6,-0.1)$) rectangle ($(L1)+(-2.4,0.1)$);
		\node at ($(L1.south)+(-1.25,-0.75)$) {$\mathbf{h}_{t-1}$};
		
		\draw[->] ($(L1.west)+(-2.5,0)$)--(L1.west) node[midway, xshift= -0.5cm, above] {$\mathbf{x}_{t}$};
		
		\draw[->] (L1.east) -- (Lout.west);
		\node at ($(Lout.east)+(1.0,\textOffset)$) {$\mathbf{y}_{t}$};
		\draw[->] (Lout.east) -- ($(Lout.east)+(1.8,0)$);
		
		\node at ($(L1.east)+(1.45,\textOffset)$) {$\mathbf{h}_{t}$};
		
		\draw[dashed] (0,-2) rectangle (6,2.5);
		\draw[dashed] (6.8,-2) rectangle (10.25,2.5);
		\node at ($(L1.north)+(0,1.25)$) {RNN-layer};
		\node at ($(Lout.north)+(0,1.25)$) {Output-layer};
	\end{tikzpicture}
	\caption{Structure of a recurrent neural network (RNN). At each time step, the input $\mathbf{x}_{t}$ and the previous hidden state $\mathbf{h}_{t-1}$ are combined and transformed by a non-linear activation function $\phi(\cdot)$ to produce the next hidden state $\mathbf{h}_{t}$. The updated state is then mapped to the output $\mathbf{y}_{t}$ via a linear output layer.}
	\label{fig:RNNStructure}
\end{figure*}
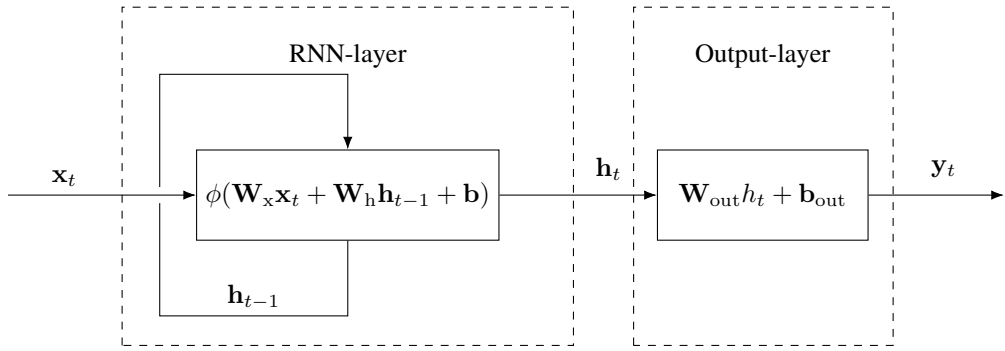

Following the hidden-state update \eqref{eq:RNNupdate}, an \emph{output layer} (typically a linear mapping)
produces the network output $\mathbf{y}_{t} \in \mathbb{R}^p$; i.e.,
\begin{equation}
	\mathbf{y}_{t} = \mathbf{W}_{\mathrm{out}} \mathbf{h}_{t} + \gls{b_out},
	\label{eq:RNNoutput}
\end{equation}
where $\mathbf{W}_{\mathrm{out}} \in \mathbb{R}^{p\times n}$ and $ \gls{b_out} \in \mathbb{R}^p$ are the output weight matrix and bias, respectively.
The overall structure of a single RNN layer paired with a linear output layer is shown in Fig.~\ref{fig:RNNStructure}.

Depending on the task, the output layer can be interpreted in two ways: 
(i) \emph{sequence-to-sequence}, where outputs are produced at every time step and stacked into a trajectory, or 
(ii) \emph{sequence-to-point}, where only a single output (often the final one) is taken as the prediction. 
Both variants share the same underlying structure, differing only in how the outputs are collected.

\subsection{Stability in RNNs}

The recurrent connections \eqref{eq:RNNupdate} that enable RNNs to retain information over long time horizons also pose challenges for training \cite{Pascanu.2013} and stability \cite{Miller.25052018}. As information is fed back through the network repeatedly, small perturbations can be progressively amplified across time, which results in exploding gradients during training~\cite{Pascanu.2013}. In such a case, the impact of noise or input perturbations on the model output does not fade or even grow with time. This leads to degraded model performance and  no horizon‑independent Lipschitz bound to exist, since $L(T)\rightarrow \infty$ as $ T\rightarrow \infty$.\\
Several methods have been proposed to enforce stability in RNNs, typically by constraining network parameters or imposing specific structural properties. Early work by Jin et al.~\cite{Liang1994RNNstability} introduced conditions for absolute stability for the fundamental RNN, while more recent approaches focus on ensuring input-to-state stability (ISS) or incremental input-to-state stability ($\delta$ISS) for various architectures, including NARX networks~\cite{BONASSI2021_NNARGS}, Echo State Networks (ESNs)~\cite{Armenio2019ModelPC}, Gated Recurrent Units (GRUs)~\cite{Bonassi_2021}, and Long Short-Term Memory (LSTM) networks~\cite{Terzi_LSTM_2019}. Such methods guarantee the \emph{existence} of a finite Lipschitz constant for a given network and therefore must be used in combination with the proposed method to ensure its applicability. In Section~\ref{subsec:Training} we discuss the training of the RNN and show how stability is ensured.

\subsection{Unrolling of RNNs}

\begin{figure*}[!t]
	\begin{center}
		\begin{tikzpicture}[actFunc/.style={rectangle, draw =black, minimum width = 2cm, minimum height = 1cm},>=Latex]
			
			\def\textOffset{0.25cm};
			\def\arrowOffset{0.25cm};
			\def\arrowLength{1.5cm};

			\node[actFunc] (L1) at (0,0) {RNN-layer};
			\node[actFunc] (L2) at  ($(L1.north)+(0,2)$) {Output-layer};

			\draw[->] (L1.east) -| ($(L1.east)+(0.5,1)$) -- ($(L1.west)+(-0.5,1)$) |- (L1.west);
			
			\draw [draw=white,fill=white] ($(L1)+(-0.1,0.9)$) rectangle ($(L1)+(0.1,1.1)$);
			\node at ($(L1.east)+(0.25,0.25)$) {$\mathbf{h}_{t}$};
			
			\draw[->] ($(L1.south)+(0,-1)$) -- ($(L1.south)$);
			\draw[->] ($(L2.north)$) -- ($(L2.north)+(0,1)$);
			\draw[->] (L1.north) -- ($(L2.south)$);
			
			\node at ($(L1.south)+(0.5,-0.5)$) {$\mathbf{x}_{t+1}$};
			\node at ($(L2.south)+(0.5,-0.5)$) {$\mathbf{h}_{t+1}$};
			\node at ($(L2.north)+(0.5,+0.5)$) {$\mathbf{y}_{t+1}$};
			
			
			\node[actFunc] (L3) at ($(L1.east)+(5,0)$) {RNN-layer};
			\node[actFunc] (L4) at  ($(L3.north)+(0,2)$) {Output-layer};
			\draw[->] ($(L3.south)+(0,-1)$) -- ($(L3.south)$);
			\draw[->] ($(L4.north)$) -- ($(L4.north)+(0,1)$);
			\draw[->] (L3.north) -- ($(L4.south)$);
			
			\node at ($(L3.south)+(0.5,-0.5)$) {$\vphantom{\mathbf{u}_{T-1}}\mathbf{x}_{1}$};
			\node at ($(L4.south)+(0.5,-0.5)$) {$\vphantom{\mathbf{h}_{T-1}}\mathbf{h}_{1}$};
			\node at ($(L4.north)+(0.5,+0.5)$) {$\vphantom{\mathbf{y}_{T-1}}\mathbf{y}_{1}$};
		
			\node[actFunc] (L5) at ($(L3.east)+(4.5,0)$) {RNN-layer};
			\node[actFunc] (L6) at  ($(L5.north)+(0,2)$) {Output-layer};
			\draw[->] ($(L5.south)+(0,-1)$) -- ($(L5.south)$);
			\draw[->] ($(L6.north)$) -- ($(L6.north)+(0,1)$);
			\draw[->] (L5.north) -- ($(L6.south)$);
			
			\node at ($(L5.south)+(0.5,-0.5)$) {$\vphantom{\mathbf{u}_{N-1}}\mathbf{x}_{N-1}$};
			\node at ($(L6.south)+(0.5,-0.5)$) {$\vphantom{\mathbf{h}_{N-1}}\mathbf{h}_{N-1}$};
			\node at ($(L6.north)+(0.5,+0.5)$) {$\vphantom{\mathbf{y}_{N-1}}\mathbf{y}_{N-1}$};

			\node[actFunc] (L7) at ($(L5.east)+(2,0)$) {RNN-layer};
			\node[actFunc] (L8) at  ($(L7.north)+(0,2)$) {Output-layer};
			\draw[->] ($(L7.south)+(0,-1)$) -- ($(L7.south)$);
			\draw[->] ($(L8.north)$) -- ($(L8.north)+(0,1)$);
			\draw[->] (L7.north) -- ($(L8.south)$);
			
			\node at ($(L7.south)+(0.5,-0.5)$) {$\mathbf{x}_{N}$};
			\node at ($(L8.south)+(0.5,-0.5)$) {$\mathbf{h}_{N}$};
			\node at ($(L8.north)+(0.5,+0.5)$) {$\mathbf{y}_{N}$};
	
			\draw[->] ($(L3.west)+(-1,0)$)--(L3.west) node[midway, above] {$\vphantom{\mathbf{h}_{t-1}}\mathbf{h}_{0}$};

			\draw[->] (L3.east)--($(L3.east)+(1.,0)$) node[midway, above] {$\vphantom{\mathbf{h}_{T-1}}\mathbf{h}_{1}$};
			\draw[->] ($(L5.west)+(-1.,0)$)--(L5.west) node[midway, above] {$\mathbf{h}_{N-1}$};
			\draw[->] (L5.east)--(L7.west) node[midway, above] {$\vphantom{\mathbf{h}_{N-1}}\mathbf{h}_{N-1}$};

			\draw[->, thick, >=Latex, line width=3pt]
			($(L1)+(2,1)$) -- ($(L3)+(-2.2,1)$)
			node[midway, above, yshift=0.2cm] {\textbf{Unrolling}};

			\node[scale = 2] at ($(L3)!0.5!(L5) + (0,.0)$) {$\mathbf{\cdots}$};
			\node[scale = 2] at ($(L4)!0.5!(L6) + (0,.0)$) {$\mathbf{\cdots}$};
		\end{tikzpicture}
	\end{center}
	
	\caption{Unfolding of a recurrent neural network (RNN) over a finite time horizon of length $N$. 
		Starting from the initial hidden state $\mathbf{h}_0$, the RNN layer and output layer are applied sequentially at each time step. 
		At each step, the past hidden state $\mathbf{h}_t$ and the current input $\mathbf{x}_{t+1}$ are combined to produce the updated hidden state $\mathbf{h}_{t+1}$ and the output $\mathbf{y}_{t+1}$. 
		The updated state is then passed to the next step and combined with the next input. This process is repeated until the final step $N$ is reached. In the unrolled representation, each time step corresponds to a distinct layer that shares parameters with all others.}
	\label{fig:RNNStructureUnrolled}
\end{figure*}
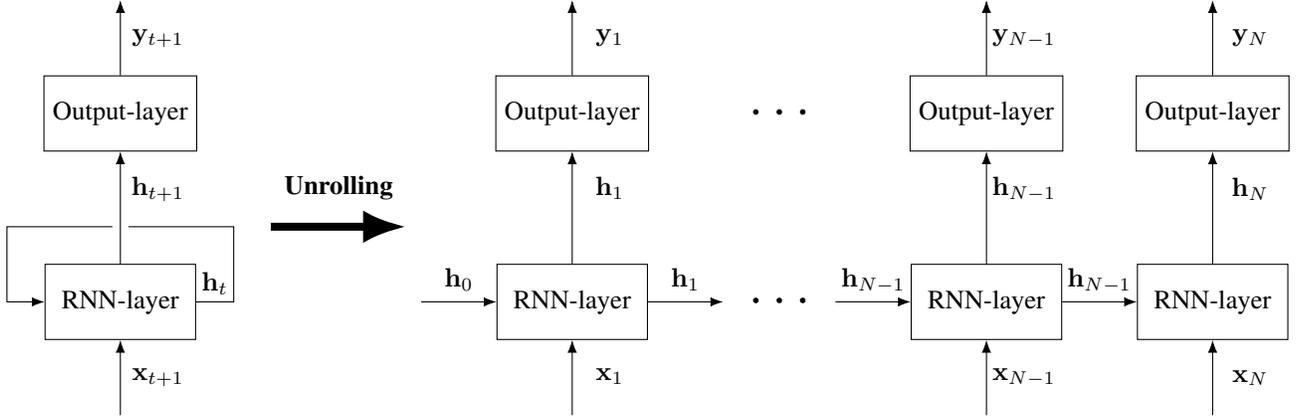

RNNs can be equivalently represented as feed-forward architectures by \emph{unrolling} the recurrence over a finite time horizon~\cite{Miller.25052018}. In this representation, each time step corresponds to a distinct layer that shares parameters with all others, this structure is shown in Fig.~\ref{fig:RNNStructureUnrolled}. For a sequence of length $N$, the hidden-state update~\eqref{eq:RNNupdate} is applied sequentially from the initial state $\mathbf{h}_0$, generating the state sequence $\mathbf{h}_t$ and the output sequence $\mathbf{y}_t$ for all $t = [1,\hdots,N]$. Unrolling a RNN removes its explicit recurrence, thereby allowing for easier analysis. The technique is therefore commonly used during the back-propagation step of the training~\cite{Miller.25052018}. In this paper, we use unrolling to track the finite‑horizon propagation of perturbations by representing the RNN by a FFNN.

\section{Methodology}
In this section is organized into four subsections. We begin in Subsection~\ref{subsec:RNN_SDP} by presenting RNN-SDP an SDP-based method for estimating RNN Lipschitz constants, inspired by the LipSDP framework for FFNNs~\cite{fazlyab2023efficientaccurateestimationlipschitz}. Building on this, in Subsection \ref{subsec:InputRestrictions} we present an approach to further tighten the bounds by incorporating known input bounds into the methodology. We then outline the data generation and stability-constrained training procedure used for evaluation in Subsection~\ref{subsec:Training}. Finally, in Subsection~\ref{subsec:EmpiricalLipschitz}, we present the design of the empirical baseline methods used for comparison.

\subsection{RNN-SDP}
\label{subsec:RNN_SDP}
To begin, we briefly recall the definition of the Lipschitz constant, as it forms the basis of the subsequent analysis. A constant \( L \) is a Lipschitz constant of a function \( f : \mathbb{R}^m \rightarrow \mathbb{R}^p \) if it satisfies the following condition\cite{Cobzas.2019}:
\begin{equation}
	\|f(\mathbf{v}_2) - f(\mathbf{v}_1)\|_2 \leq L \|\mathbf{v}_2 - \mathbf{v}_1\|_2 \quad \forall \mathbf{v}_1, \mathbf{v}_2\in\mathbb{R}^m
	\label{eq:LipschitzGeneral}
\end{equation}
Since the inequality~\eqref{eq:LipschitzGeneral} holds for all \( \mathbf{v}_1, \mathbf{v}_2 \in \mathbb{R}^m \), we can guarantee that the function $f(\cdot)$ does not amplify input differences by more than a factor of \( L \). In the context of NNs, accurate knowledge of \( L \) enables us to estimate the system's sensitivity to input noise and other perturbations. This is particularly valuable, as neural networks are often highly non-linear, making it difficult to reason about their behavior without such bounds~\cite{Liu.2021}.\\

Inspired by \cite{fazlyab2023efficientaccurateestimationlipschitz}, we interpret the function $f(\cdot)$ as a neural network, transforming a given input $\mathbf{v}_i$ into the network-output $f(\mathbf{v}_i)$. For an RNN with the update and output equations \eqref{eq:RNNupdate},\eqref{eq:RNNoutput}, the input can be defined as:
\begin{equation}
	\mathbf{v}_i = 
	\begin{pmatrix}
		\gls{x}\\
		\mathbf{h}_0
	\end{pmatrix}_i,
	\label{eq:inputdef}
\end{equation}
with $\mathbf{h}_0 \in \mathbb{R}^n$ being the initialization of the hidden state and $\gls{x} = [\mathbf{x}^{T}_1,\mathbf{x}^{T}_2,\dots,\mathbf{x}^{T}_N]^T \in \mathbb{R}^{mN}$ as a stack of all input vectors $\mathbf{x}_t$ at all $N$ discrete time steps of the sequence. Depending on if the RNN makes point-wise or sequential predictions (see Section \ref{subsec:RNNDefinition}) the network output is either the output at the last time step $\mathbf{y}_N  \in \mathbb{R}^{p}$ or the full sequence of outputs $\gls{y} = [\mathbf{y}^T_1,\mathbf{y}^T_2,\dots,\mathbf{y}^T_N]^T \in \mathbb{R}^{pN}$. For this application we choose a network with a point-wise output at the last time step, the reasoning for this choice is discussed in Appendix \ref{ap:SeqVsPoint}.
Using \eqref{eq:inputdef} We can therefore write \eqref{eq:LipschitzGeneral} for \eqref{eq:RNNupdate}, \eqref{eq:RNNoutput} as:
\begin{equation}
	||\mathbf{y}_{N,2}-\mathbf{y}_{N,1}||_2 
	-L 
	\left\lVert
	\begin{pmatrix} \gls{x}\\\mathbf{h}_0	\end{pmatrix}_2
	-
	\begin{pmatrix} \gls{x}\\\mathbf{h}_0	\end{pmatrix}_1
	\right\rVert_2 \leq  0.
	\label{eq:LispschitzRNN}
\end{equation}
For any two input vectors $v_1 = [\gls{x}^T,\mathbf{h}_0^T]^T_1$ and $v_2 = [\gls{x}^T,\mathbf{h}_0^T]^T_2$ and their corresponding output vectors $\mathbf{y}_{N,2}$ and $\mathbf{y}_{N,1}$, we can be related to the internal states of the RNN. As such we define a joined state vector:
\begin{equation}
	\gls{z} = [\gls{x}^T,\gls{h}^T]^T \in \mathbb{R}^{d_z}, \quad d_z = mN+n(N+1),
\end{equation} 
with $\gls{h} = [\mathbf{h}_0^T,\mathbf{h}_1^T,\hdots,\mathbf{h}_N^T]^T \in \mathbb{R}^{n(N+1)}$ being the stacked vector of all internal states of the unrolled RNN. The definition of \gls{z} allows us to write the input, internal states and output for \eqref{eq:RNNupdate}, \eqref{eq:RNNoutput} as:
\begin{equation}
	\begin{aligned}
			\begin{pmatrix}
				\gls{x}\\ 
				\mathbf{h}_0
			\end{pmatrix} =& \mathbf{E}_\mathrm{in} \gls{z},\\
			\mathbf{y}_N =& \mathbf{E}_\mathrm{out} \gls{z} + \gls{b_out},
	\end{aligned}
	\label{eq:zRelations}
\end{equation}
with
\begin{equation}
	\begin{aligned}
		\mathbf{E}_\mathrm{in}& = 
		\begin{pmatrix}
				\mathbf{I}_{Nm}& \gls{0} &\gls{0}\\
				 \gls{0}& \mathbf{I}_n &\gls{0}\\
				 \gls{0} &\gls{0}&\gls{0}
		\end{pmatrix} \in \mathbb{R}^{(mN+n) \times d_z},
		\\
		\mathbf{E}_\mathrm{out} &= \begin{pmatrix}
			\gls{0} &\gls{W_out}
		\end{pmatrix}\in \mathbb{R}^{p \times d_z}.
	\end{aligned}
\end{equation}
By squaring \eqref{eq:LispschitzRNN} and inserting \eqref{eq:zRelations}, we get:

\begin{equation}
	||\mathbf{E}_\mathrm{out}(\gls{z}_2-\gls{z}_1)||_2^2
	-L^2 
	||
		\mathbf{E}_\mathrm{in}
	(\gls{z}_2-\gls{z}_1)	
	||_2^2 \leq  0.
	\label{eq:LispschitzSubstituted}
\end{equation}
Rewriting the squared $l_2$-norms as inner products and introducing $\rho= L^2$ yields:
\begin{equation}
	\begin{aligned}
		(\gls{z}_2-\gls{z}_1)^T
		\mathbf{E}_\mathrm{out}^T
		\mathbf{E}_\mathrm{out} (\gls{z}_2-\gls{z}_1) \qquad\quad\qquad\\
		-\rho
		(\gls{z}_2-\gls{z}_1)^T\mathbf{E}_\mathrm{in}^T\mathbf{E}_\mathrm{in} (\gls{z}_2-\gls{z}_1) \le 0.
	\end{aligned}
	\label{eq:LipschitzInnerProduct}
\end{equation}
Since both inner products in \eqref{eq:LipschitzInnerProduct} are of the same dimension, the matrices can be joint into a singular matrix $\mathbf{M} \in \mathbb{R}^{d_z\times d_z}$:
\begin{equation}
	(\gls{z}_2-\gls{z}_1)^T \mathbf{M} (\gls{z}_2-\gls{z}_1) \leq 0,
	\label{eq:LipschitzM}
\end{equation}
with:
\begin{equation}
	\begin{aligned}
	\mathbf{M}&= \mathbf{E}_{out}^T\mathbf{E}_{out}-\rho\mathbf{E}_{in}^T\mathbf{E}_{in}
	\\ &=
	\begin{pmatrix}
		-\rho \,\mathbf{I}_{Nm} & \gls{0}&\gls{0}&\gls{0}\\
		\gls{0}&-\rho \,\mathbf{I}_n&\gls{0}&\gls{0}\\
		\gls{0}&\gls{0}& \gls{0} &\gls{0} \\
		\gls{0} &\gls{0} &\gls{0}& \gls{W_out}^T\gls{W_out} \\
	\end{pmatrix} .
	\end{aligned}
\end{equation}
The inequality~\eqref{eq:LipschitzM} would hold for all \( \gls{z}_1, \gls{z}_2 \in \mathbb{R}^{d_z} \) if and only if \( \mathbf{M} \preceq 0 \). However, this condition is only satisfied in the trivial case, where \( \gls{W_out} = \gls{0} \), implying that the neural network outputs zero regardless of the input. In this scenario, any \( \rho > 0 \) trivially satisfies the Lipschitz condition. For any non-zero output weights, the Gram matrix \( \gls{W_out}^T \gls{W_out} \) is positive definite. Consequently, since the upper-left block of \( \mathbf{M} \) is negative definite and the lower-right block is positive definite, the matrix \( \mathbf{M} \) is indefinite, see also \cite{fazlyab2023efficientaccurateestimationlipschitz}.

It is important to note, however, that the the internal states \( \gls{z}_1, \gls{z}_2 \), while of dimension \( d_z \), are not arbitrary vectors in \( \mathbb{R}^{d_z} \). Due to the dependencies induced by the layer-wise interactions within the network, these individual vectors making up \( \gls{z} \) cannot be freely chosen, restricting the valid-values of \( \gls{z}_1, \gls{z}_2 \) to a subset \( \mathcal{Z} \subseteq \mathbb{R}^{d_z} \). In the following, we introduce supplementary equations that reflect the interactions and constraints defining \( \mathcal{Z} \), in order to derive an expression guaranteeing the satisfaction of the inequality in~\eqref{eq:LipschitzM}.

We begin by examining the non-linear activation functions within NNs. For all commonly used NN activation functions, the non-linearity \( \varphi(\cdot) \) is \textit{slope-restricted}~\cite{fazlyab2021safetyverificationrobustnessanalysis}. This includes the $\tanh(\cdot)$ non-linearity used in the RNN update step \eqref{eq:RNNupdate}. We denote a generic slope-restricted function as \( \varphi(v) \) if applied to a scalar $v$ and as $\phi(\mathbf{v}) = [\varphi(v_1),\varphi(v_2),\dots,\varphi(v_n)]^T,\,\phi: \mathbb{R}^n\rightarrow \mathbb{R}^n$ if applied element wise to a vector $\mathbf{v}$. For \( \varphi(\cdot) \) the following inequality holds for any two inputs \( v_1, v_2 \in \mathbb{R} \)\cite{fazlyab2023efficientaccurateestimationlipschitz}:

\begin{equation}
	\alpha \leq \frac{\varphi(v_2)-\varphi(v_1)}{v_2-v_1} \leq \beta  , 
	\label{eq:slopeBoundGeneral}
\end{equation}
where \( \alpha \) and \( \beta \) define the minimum and maximum slopes, respectively. This condition can equivalently be expressed as:
\begin{equation}
	\left( \frac{\varphi(v_2) - \varphi(v_1)}{v_2 - v_1} - \alpha \right)\left( \frac{\varphi(v_2) - \varphi(v_1)}{v_2 - v_1} - \beta \right) \leq 0,
\end{equation}
which, after clearing the denominator, becomes:
\begin{equation}
	\begin{aligned}
		\left( \varphi(v_2) - \varphi(v_1) - \alpha(v_2 - v_1) \right)\qquad\qquad\qquad
		\\
		\left( \varphi(v_2) - \varphi(v_1) - \beta(v_2 - v_1) \right) \leq 0.
	\end{aligned}
	\label{eq:scalarInequ}
\end{equation}
As shown in \cite{fazlyab2023efficientaccurateestimationlipschitz}, the scalar inequality~\eqref{eq:scalarInequ} can be rewritten in quadratic form. Multiplying both sides by $-2$ to eliminate the fractional coefficients yields:
\begin{equation}
	\begin{aligned}
	\begin{pmatrix}	v_2-v_1 \\ \!\varphi(v_2)\!-\!\varphi(v_1)\!	\end{pmatrix}^{\!\!T} 
	\!\!\!\!\begin{pmatrix}	-2 \alpha \beta  & (\alpha\!+\!\beta)\!\\\!(\alpha\!+\!\beta) & -2 	\end{pmatrix}
	 \!\!\!\begin{pmatrix}	v_2-v_1\\\! \varphi(v_2)\!-\!\varphi(v_1)\!	\end{pmatrix} \!\!\geq \!0
	 \end{aligned}.
	 \label{eq:ScalarSlopConstraint}
\end{equation}
To extend \eqref{eq:ScalarSlopConstraint} to vector-valued functions, we use the element-wise non-linear transformation $\phi(\mathbf{v})$ as well as the set
\begin{equation}
	\mathcal{T}_n = \left\{\gls{T} \in \mathbb{S}^n \mid \gls{T} = diag(\lambda)|\lambda_{ii} \in \mathbb{R}^n_{\geq0} \right\},
	\label{eq:Tdefinition}
\end{equation}
where $\mathbb{S}^n$ denotes the set of symmetric $n\times n$ matrices. Then, by Sylvester's law of inertia \cite{Sylvester.1852}, for any $\gls{T} \in \mathcal{T}_n$ the following inequality holds for all $ \mathbf{v}_1,\mathbf{v}_2 \in \mathbb{R}^n$:

\begin{equation}
	\begin{pmatrix}	\mathbf{v}_2 \shortminus \mathbf{v}_1 \\ \!\phi(\mathbf{v}_2)\shortminus \phi(\mathbf{v}_1)\!\!	\end{pmatrix}^{\!\!\! T} 
	\!\!\!\!
	\begin{pmatrix}	\shortminus 2 \alpha \beta \gls{T} & (\alpha \shortplus \beta)\gls{T}\! \\ \!\!(\alpha\shortplus\beta)\gls{T} & \shortminus 2 \gls{T}	\end{pmatrix} 
	\!\!\!
	\begin{pmatrix}	\mathbf{v}_2\shortminus \mathbf{v}_1\\ \! \phi(\mathbf{v}_2)\shortminus\phi(\mathbf{v}_1)\!\!	\end{pmatrix} 
	\!\!\geq\! 0,
	\label{eq:FazIneq}
\end{equation}
By viewing $\mathbf{v}_i,\,i\in \{1,2\}$ as the linearly transformed inputs into each network layer and $\varphi(\mathbf{v}_i)$ as the layer output after the non-linear activation function has been applied, the slope restrictions \eqref{eq:ScalarSlopConstraint} can be applied to the network as a whole~\cite{fazlyab2023efficientaccurateestimationlipschitz}. To apply this thinking in the RNN setting, we need to reformulate the update function~\eqref{eq:RNNupdate} to apply to the entire network instead of just one time step. Using the stacked inputs \gls{x} and hidden states \gls{h} we obtain:

\begin{equation}
	\mathbf{B}_\mathrm{h}\gls{h} = \phi(\mathbf{A}_\mathrm{h}\gls{h}+\textbf{A}_\mathrm{x}\gls{x} + \mathbf{\tilde{b}}),
	\label{eq:NetworkUpdateFunction}
\end{equation}
with:

\begin{equation}
	\begin{aligned}
		\mathbf{A}_\mathrm{x} &= 
		\begin{pmatrix}
			\gls{W_x} &   & \gls{0}\\
			 & \ddots &\\
			\gls{0}&  &\gls{W_x}
		\end{pmatrix}
		\!,\;
		\mathbf{A}_\mathrm{h} = 
		\begin{pmatrix}
			\gls{W_h} &   & \gls{0} &\gls{0}\\
			& \ddots && \vdots\\
			\gls{0}&  &\gls{W_h}&\gls{0}
		\end{pmatrix},
		\\
		\mathbf{B}_\mathrm{h} &= 
		\begin{pmatrix}
			 \gls{0} & \mathbf{I}_{n} & &\gls{0} \\
			\vdots & &\ddots & \\
			 \gls{0}& \gls{0}&  &\mathbf{I}_{n}
		\end{pmatrix}
		\!,\;
		\tilde{\mathbf{b}} =
		\begin{pmatrix} \mathbf{b}^T,\hdots, \mathbf{b}^T		\end{pmatrix}^T.
	\end{aligned}
\end{equation}
Now we unite \gls{x} and \gls{h} as the joined state vector  \gls{z} and can rewrite \eqref{eq:NetworkUpdateFunction} compactly as:
\begin{equation}
	\mathbf{B}
	\gls{z} 
	=
	\varphi(
	\mathbf{A}
	\gls{z}
	+
	\tilde{\mathbf{b}}
	),
	\label{eq:JointStateUpdateFunction}
\end{equation}
with:
\begin{equation}
	\mathbf{B}
	=
	\begin{pmatrix}
		\gls{0} & \mathbf{B}_\mathrm{h}
	\end{pmatrix}
	,\qquad
	\mathbf{A}
	=
	\begin{pmatrix}
		\mathbf{A}_\mathrm{x} & \mathbf{A}_\mathrm{h}
	\end{pmatrix}.
\end{equation}
Using the established layer relations, we substitute layer input $\mathbf{v}_i = \mathbf{A}\gls{z}+\tilde{\mathbf{b}}$ in~\eqref{eq:FazIneq}. From \eqref{eq:JointStateUpdateFunction} it follows that the layer output therefore is $\phi(\mathbf{v}_i)= \mathbf{B}\gls{z}$. Applying these substitutions to~\eqref{eq:FazIneq}, we get:

\begin{equation}
	\begin{pmatrix}	
		\! \mathbf{A}(\gls{z}_2\shortminus\gls{z}_1)\! \\ 
		\! \mathbf{B}(\gls{z}_2\shortminus\gls{z}_1)\!
	\end{pmatrix}^{\!\! T} 
	\!\!\!
		\begin{pmatrix}	
			\shortminus2 \alpha \beta \gls{T} & (\alpha\shortplus\beta)\gls{T} \\
			(\alpha\shortplus\beta)\gls{T} & \shortminus2 \gls{T}	
		\end{pmatrix} 
	\!\!
	\begin{pmatrix}	
		\! \mathbf{A}(\gls{z}_2\shortminus\gls{z}_1)\! \\ 
		\! \mathbf{B}(\gls{z}_2\shortminus\gls{z}_1)\!
	\end{pmatrix} 
	\geq 0 .
\end{equation}
By factoring out $(\gls{z}_2-\gls{z}_1)$ we obtain the following equation:

\begin{equation}
	(\gls{z}_2 - \gls{z}_1)^T
	\mathbf{Q}
	(\gls{z}_2 - \gls{z}_1) \geq 0,
	\label{eq:ConstructionQ}
\end{equation}
where $\mathbf{Q}\in \mathbb{R}^{d_z\times d_z}$ is defined as :
\begin{equation}
	\mathbf{Q}=
	\begin{pmatrix}
		\mathbf{A} \
		\mathbf{B}
	\end{pmatrix}^T
	\begin{pmatrix}
		-2\alpha\beta \gls{T} & (\alpha + \beta)\gls{T} \\
		(\alpha + \beta)\gls{T} & -2\gls{T}
	\end{pmatrix}
	\begin{pmatrix}
		\mathbf{A} \
		\mathbf{B}
	\end{pmatrix}.
	\label{eq:Qdef}
\end{equation}
Combining the inequalities \eqref{eq:ConstructionQ} and \eqref{eq:LipschitzM} yields:

\begin{equation}
	(\gls{x}_2\shortminus\gls{x}_1)	^T
	(\mathbf{M}\shortplus\mathbf{Q})
	(\gls{x}_2\shortminus\gls{x}_1)	
	\geq 
	(\gls{x}_2\shortminus\gls{x}_1)	^T
	(\mathbf{M})
	(\gls{x}_2\shortminus\gls{x}_1).
	\label{eq:joinedInequalities}
\end{equation}
If \( (\mathbf{Q} + \mathbf{M}) \preceq 0 \), then the left-hand side of the inequality \eqref{eq:joinedInequalities} is guaranteed to be less than or equal to zero. This in turn also guarantees the right hand side to be less or equal to zero. This implies that~\eqref{eq:FazIneq} is satisfied, meaning any value of \( \rho \) for which \( (\mathbf{Q} + \mathbf{M}) \) is negative semi-definite constitutes a valid upper bound on the Lipschitz constant. Consequently, the Lipschitz constant can be estimated by solving the following optimization problem:

\begin{equation}
	\min_{\rho,\mathbf{T}} \rho \qquad \text{s.t. } (\mathbf{M}+\mathbf{Q}\preceq 0)
	\label{eq:FazlyapOptim}
\end{equation}
The decision variables in this optimization problem are \( \rho \in \mathbb{R}^+ \) and \( \mathbf{T} \in \mathcal{T}_n \), where \( \mathcal{T}_n \) in \eqref{eq:Tdefinition} is a convex set. Since the matrix \( (\mathbf{M} + \mathbf{Q}) \) is affine in both \( \rho \) and \( T \), the problem is a semidefinite program (SDP) and can be solved efficiently using numerical solvers \cite{GartnerBerndandMatousekJiri.2012}. Solving the convex formulation yields a global optimum $\rho$, which gives an upper bound on the squared Lipschitz constant of the unrolled RNN, yielding \( L = \sqrt{\rho} \).

However, \eqref{eq:FazlyapOptim} only gives the Lipschitz bound for the last output of the sequence, since we chose the network output to be point-wise, see \eqref{eq:LispschitzRNN}. Therefore, \eqref{eq:FazlyapOptim} does not provide a bound for the deviations that could occur during the sequence, but only at the end. To get a bound for the entire sequence, the optimization~\eqref{eq:FazlyapOptim} needs to be repeated $N$ times, once for each possible sequence length. The overall bound is then obtained by taking:
\begin{equation}
	\begin{aligned}
	\max_i \left(\min_{\rho,\mathbf{T}} \rho \qquad \text{s.t. } (\mathbf{M}+\mathbf{Q}\preceq 0)\right),\\
	\text{with } i\in [1,2\hdots,N].
\end{aligned}
	\label{eq:FinalOptim}
\end{equation}

\subsection{Input Restrictions}
\label{subsec:InputRestrictions}
\begin{figure}[t]
	\centering
	\subfloat[]{%
		\begin{tikzpicture}[>=Latex]
			\begin{axis}[
				name = topaxis,
				width = 0.65\linewidth,
				height = 4cm,
				axis lines=middle,
				xmin=-3, xmax=3,
				ymin=-1.5, ymax=1.5,
				ticks=none,
				domain=-3:3, samples=200
				]
				\node[anchor = west] at (-2.75,1.25){$\tanh(v)$};
				\addplot[very thick,red] {tanh(x)};
			\end{axis}

			\begin{axis}[
				at = {(topaxis.south west)},
				anchor = north west,
				name = bottomaxis,
				yshift = -0.5cm,
				width = 0.65\linewidth,
				height = 3.5cm,
				axis lines=middle,
				xmin=-3, xmax=3,
				ymin=-0.25, ymax=1.5,
				ticks=none,
				domain=-3:3, samples=200
				]
				\node[anchor = west] at (-2.75,1.25){$\frac{1}{\cosh(v)^2}$};
				\addplot[very thick,red] {1/cosh(x)^2};
				\addplot[only marks, mark=* , mark size=2pt, black] coordinates {(0, 1)} node[above right] {$\beta = 1$};
				\node(alpha) at (1,0.3) [above right] {$\alpha = 0$};
				\node(alpha2) at (-1,0.3) [above left]  {$\alpha = 0$};
			\end{axis}
			\draw[->] (alpha.east)-- ++(0.35,0);
			\draw[->] (alpha2.west)-- ++(-0.35,0);
		\end{tikzpicture}%
	}\hfill
	\subfloat[]{%
		\begin{tikzpicture}[>=Latex]
			\begin{axis}[
				name = topaxis,
				width = 0.65\linewidth,
				height = 4cm,
				axis lines=middle,
				xmin=-3, xmax=3,
				ymin=-1.5, ymax=1.5,
				ticks=none,
				domain=-3:3, samples=200
				]
				\addplot[very thick,red] {tanh(x)};
				\addplot[dashed,blue]coordinates {(1,-1.25) (1,1.25)}node[right]{ub};
				\addplot[dashed,blue]coordinates {(-1,-1.25) (-1,1.25)}node[left]{lb};
			\end{axis}

			\begin{axis}[
				at = {(topaxis.south west)},
				anchor = north west,
				name = bottomaxis,
				yshift = -0.5cm,
				width = 0.65\linewidth,
				height = 3.5cm,
				axis lines=middle,
				xmin=-3, xmax=3,
				ymin=-0.25, ymax=1.5,
				ticks=none,
				domain=-3:3, samples=200
				]
				\addplot[very thick,red] {1/cosh(x)^2};
				\addplot[dashed,blue]coordinates {(1,-1.25) (1,1.25)};
				\addplot[dashed,blue]coordinates {(-1,-1.25) (-1,1.25)};
				\addplot[only marks, mark=* , mark size=2pt, black] coordinates {(0, 1)} node[above right] {$\beta = 1$};
				\addplot[only marks, mark=* , mark size=2pt, black] coordinates {(1, {1/cosh(1)^2})} node[above right,xshift = -2pt] {$\alpha \approx 0.42$};
				\addplot[only marks, mark=* , mark size=2pt, black] coordinates {(-1, {1/cosh(1)^2})} node[above left,xshift = 2pt] {$\alpha \approx 0.42$};
			\end{axis}
		\end{tikzpicture}%
	}
	\caption{Impact of upper bound (ub) and lower bound (lb) on the slope restriction parameters $\alpha$ and $\beta$. (a) unbounded (b) bounded by (lb,ub) = (-1,1)}
	\label{fig:SlopeRestrictions}
\end{figure}
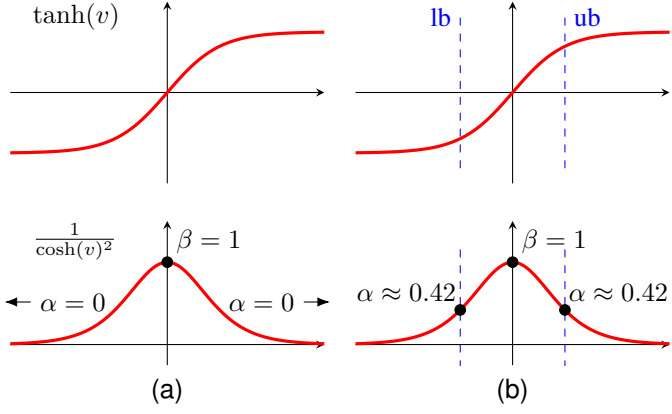

Using global slope bounds $(\alpha, \beta)$ as in \eqref{eq:slopeBoundGeneral} to estimate the Lipschitz constant yields a valid bound for any possible input. In practice, however, NN inputs are often normalized to a fixed range, improving both training convergence and numerical stability~\cite{Fergus2022}. This normalization effectively constrains the inputs of the NN, which in turn can be used to derive tighter local slope bounds, as illustrated in Fig.~\ref{fig:SlopeRestrictions}. In this section we explore an approach to exploit this observation to compute \emph{local slope bounds}, with the aim to further tighten the Lipschitz constant estimates.

We illustrate the proposed approach using the hyperbolic tangent activation function  $f: R^1\rightarrow (-1,1)$,
\begin{equation}
	\varphi(v) = \tanh(v), \qquad \dot{\varphi}(v) = \frac{1}{\cosh^2(v)}.
	\label{eq:activationFunctionDefinition}
\end{equation}
The derivative in~\eqref{eq:activationFunctionDefinition} is symmetric about $v = 0$ and decreases monotonically with $|v|$. The maximum slope $\beta$ of $\varphi$ therefore occurs at $v = 0$ and decreases as $|v|$ grows. Given the pre-activation bounds $[v_{\mathrm{lb},i}^{(l)}, v_{\mathrm{ub},i}^{(l)}]$ of a neuron $i$ in layer $l$, the local slope bounds follow directly from evaluating $\dot{\varphi}(v)$ at the extreme points of this interval:
\begin{equation}
	\begin{aligned}
	\alpha^{(l)}_i &= \frac{1}{\cosh^2\!\left( \max(|v^{(l)}_{\mathrm{lb},i}|, |v^{(l)}_{\mathrm{ub},i}|) \right)},\\
	\beta^{(l)}_i &=
	\begin{cases}
		1 & 0 \in [v^{(l)}_{\mathrm{lb},i}, v^{(l)}_{\mathrm{ub},i}] \\
		\frac{1}{\cosh^2\!\left( \min(|v^{(l)}_{\mathrm{lb},i}|, |v^{(l)}_{\mathrm{ub},i}|) \right)} & \text{otherwise}
	\end{cases}.
	\end{aligned}
	\label{eq:localslope}
\end{equation}
For other differentiable slope-restricted activation functions, the same principle applies, although determining the extrema may not be as direct as in the $\tanh(\cdot)$ case. The pre-activation bounds $[v^{(l)}_{\mathrm{lb},i}, v^{(l)}_{\mathrm{ub},i}]$ depend on bounds on the current input $\mathbf{u}$ and the previous hidden state $\mathbf{h}^{(l-1)}$ as well as the bias $\mathbf{b}$. Input bounds $[\mathbf{u}_{\mathrm{lb}}, \mathbf{u}_{\mathrm{ub}}]$ are assumed to be constant and known based on normalization applied to the inputs of the NN. The hidden-state bounds $[\mathbf{h}^{(l)}_{\mathrm{lb}}, \mathbf{h}^{(l)}_{\mathrm{ub}}]$ evolve layer by layer: the bounds of the initial state $\mathbf{h}^{(0)}$ are conservatively set to $[-1,1]$ element-wise, matching the output range of $\tanh(\cdot)$. For $l > 0$, bounds are updated from the previous layer’s post-activation outputs. 
The pre-activation limits are obtained by solving:
\begin{equation}
	\begin{aligned}
		v_{\mathrm{lb},i}^{(l)} &= \min_{\mathbf{x},\, \mathbf{h}} \; \gls{W_x}_{,i} \mathbf{x} + \gls{W_h}_{,i} \mathbf{h} + \mathbf{b}_i, \\
		v_{\mathrm{ub},i}^{(l)} &= \max_{\mathbf{x},\, \mathbf{h}} \; \gls{W_x}_{,i} \mathbf{x} + \gls{W_h}_{,i} \mathbf{h} + \mathbf{b}_i, \\
		&\text{s.t. } \mathbf{x} \in [\mathbf{x}_{\mathrm{lb}}, \mathbf{x}_{\mathrm{b}}],\;
		\mathbf{h} \in [\mathbf{h}^{(l-1)}_{\mathrm{lb}}, \mathbf{h}^{(l-1)}_{\mathrm{ub}}],
	\end{aligned}
	\label{eq:inputbound}
\end{equation}
and can then be used to update the hidden-state bounds for layer $l$:
\begin{equation}
	\begin{aligned}
		h_{\mathrm{lb},i}^{(l)} &= \tanh\!\left(v_{\mathrm{lb},i}^{(l)}\right), \\
		h_{\mathrm{ub},i}^{(l)} &= \tanh\!\left(v_{\mathrm{ub},i}^{(l)}\right).
	\end{aligned}
	\label{eq:nexthbound}
\end{equation}
 These updated bounds $[\mathbf{h}^{(l)}_{\mathrm{lb}}, \mathbf{h}^{(l)}_{\mathrm{ub}}]$ are then used in~\eqref{eq:inputbound} for the following layer. Repeating the calculations in~\eqref{eq:inputbound}, \eqref{eq:localslope}, and~\eqref{eq:nexthbound}  as shown in Algorithm~\ref{alg:slope_bounds} for all layers of the NN yields the complete set of local slope bounds $\alpha^{(l)}_i$ and $\beta^{(l)}_i$ for each neuron~$i$ in each layer~$l$. 

\begin{algorithm}[H]
	\caption{Local Slope Bound Estimation}
	\label{alg:slope_bounds}
	\begin{algorithmic}[1]
		\STATE \textbf{Input:} Input and initial hidden bounds $[x_{lb}, x_{ub}]$, $[h^0_{lb}, h^0_{ub}]$
		\STATE \textbf{Initialize:} $l \gets 0$
		\REPEAT
		\STATE Compute input bounds $[v^{(l)}_{lb,i}, v^{(l)}_{ub,i}]$ using~\eqref{eq:inputbound}
		\STATE Compute local slope bounds $[\alpha^{(l)}_i, \beta^{(l)}_i]$ using~\eqref{eq:localslope}
		\STATE Compute next layer bounds $[\mathbf{h}_{lb}^{l+1}, \mathbf{h}_{ub}^{l+1}]$ using~\eqref{eq:nexthbound}
		\STATE $l \gets l + 1$
		\UNTIL{$l$ is the last layer}
		\STATE \textbf{Output:} All local slope bounds $\alpha^{(l)}_i$, $\beta^{(l)}_i$
	\end{algorithmic}
\end{algorithm}

In the original formulation of the $\mathbf{Q}$ matrix in~\eqref{eq:Qdef}, the slope-restriction parameters $\alpha$ and $\beta$ were assumed identical for all neurons in a layer. This uniformity does not hold once neuron-specific local bounds $\alpha^{(l)}_i$ and $\beta^{(l)}_i$ have been computed. To accommodate per-neuron parameters at layer $l$, replace the scalars in~\eqref{eq:Qdef} with diagonal matrices
\begin{equation}
	\begin{aligned}
		\mathbf{D}_{\times} &= \operatorname{diag}\!\big(
		\alpha^{(1)}_1\beta^{(1)}_1, \ldots,
		\alpha^{(1)}_n\beta^{(1)}_n,\\ 
		&\qquad\qquad\qquad\alpha^{(2)}_1\beta^{(2)}_1, \ldots, 
		\alpha^{(l)}_{n}\beta^{(l)}_{n}\big),
	 \\
	\mathbf{D}_{+} &= \operatorname{diag}\!\big(
	\alpha^{(1)}_1{+}\beta^{(1)}_1, \ldots,
	\alpha^{(1)}_n{+}\beta^{(1)}_n,\\ 
	&\qquad\qquad\qquad\alpha^{(2)}_1{+}\beta^{(2)}_1, \ldots, 
	\alpha^{(N)}_{n}{+}\beta^{(N)}_{n}\big),
	\end{aligned}
\end{equation}
and define
\begin{equation}
	\mathbf{Q} =
	\begin{pmatrix} \mathbf{A} & \mathbf{B} \end{pmatrix}^{\!\top}
	\begin{pmatrix}
		-2\,\mathbf{D}_{\times}\,\gls{T} & \;\mathbf{D}_{+}\,\gls{T} \\
		\;\mathbf{D}_{+}\,\gls{T} & -2\,\gls{T}
	\end{pmatrix}
	\begin{pmatrix} \mathbf{A} & \mathbf{B} \end{pmatrix}.
	\label{eq:QLocalSlope}
\end{equation}
When $\alpha^{(l)}_i \equiv \alpha$ and $\beta^{(l)}_i \equiv \beta$ for all $i$ and $l$, we have $\mathbf{D}_{\times} = (\alpha\beta)\,\mathbf{I}$ and $\mathbf{D}_{+} = (\alpha{+}\beta)\,\mathbf{I}$, and~\eqref{eq:QLocalSlope} reduces to the global-slope form in~\eqref{eq:Qdef}.

The formulation in \eqref{eq:QLocalSlope} preserves the structural role of $\mathbf{Q}$, enabling the main method to be applied without major changes, while accommodating neuron-specific slope constraints. The impact of these modifications on the Lipschitz constant estimation is examined in Section~\ref{sec:Results}.

\subsection{Network Training}
\label{subsec:Training}
Appropriate training is essential for achieving good performance in any NN. For RNNs, however, special care must be taken to ensure \emph{stability},  
as this guarantees a finite Lipschitz constant and allows our method to be applied. The primary component of our employed loss function is the standard mean squared error (MSE) on the RNN output:  
\begin{equation}
	\mathcal{L}_{\mathrm{ac}} = 
	\frac{1}{N-25}\sum_{i = 25}^{N}(\mathbf{y}_i - \mathbf{y}_i^*)^2 ,
\end{equation}
where the first 25 samples of each sequence are omitted (wash-out) to avoid learning from  
initial transients that do not reflect the true system dynamics\cite{Jaeger.2001}.  

To promote stability, we include a regularization term that activates when the spectral norm of the recurrent weight matrix $\mathbf{W}_h$ exceeds~1, since $\|\mathbf{W}_h\|_2 < 1$ is the condition for absolute stability in RNNs~\cite{Liang1994RNNstability}.  
This term is implemented as:

\begin{equation}
	\mathcal{L}_{\mathrm{st}} = \operatorname{LReLU}\!\big(\|\mathbf{W}_h\|_2 - 1\big),
\end{equation}
with the leaky rectified linear unit
\begin{equation}
	\operatorname{LReLU}(x) =
	\begin{cases}
		a_1 x, & x \ge 0, \\
		a_2 x, & x < 0,
	\end{cases}
\end{equation}
and $a_1 \gg a_2$.  
This choice of $a_1$ and $a_2$ strongly penalizes $\|\mathbf{W}_h\|_2 > 1$ while making further reductions below one  
have only a minor effect, thus avoiding unnecessary loss of accuracy. 

The full training objective is therefore
\begin{equation}
	\mathcal{L} = \mathcal{L}_{\mathrm{ac}} + \mathcal{L}_{\mathrm{st}} .
\end{equation}
RNNs were trained using the Adam optimizer\cite{Kingma.2017}. Training stopped if the validation loss  
did not improve for ten consecutive epochs and $\|\mathbf{W}_h\|_2 < 1$ was satisfied.

The training data for the RNNs was synthetically generated using a multi-tank system. The input at each time step $t$ is defined as $\mathbf{x}_t = [u_0,u_1,\ldots,u_n]^\top$, where $u_i$ denotes the inflow into tank $i$. The corresponding network output is $\mathbf{y}_t = [h_0,h_1,\ldots,h_n]^\top$, representing the tank water levels. The tank dynamics are modeled in discrete time with step size $\Delta t = 1\,\text{s}$, based on the simplified physical approximation:  
\begin{equation}
	h_{t+1,i}\!=\! 
	\begin{cases}
		\! h_{t,i} \shortplus \Delta t \big(u_{t,i} \shortminus a_i\sqrt{h_{t,i}}\big), 										&\!\!\!\! i\!=\!0 \\[6pt]
		\! h_{t,i} \shortplus \Delta t \big(u_{t,i} \shortplus a_{i-1}\sqrt{h_{t,i-1}} \shortminus a_i\sqrt{h_{t,i}}\big), &\!\!\!\! i \! >\!0
	\end{cases},
	\label{eq:Multitank}
\end{equation}
where $a_i$ is a positive parameter reflecting the outflow characteristics of tank $i$ (e.g., pipe dimensions). Using~\eqref{eq:Multitank}, synthetic data for a three-tank setup was generated. The resulting data set consists of 1000 sequences of length 100, split into 70\% for training and 30\% for validation.

\subsection{Empirical Lipschitz Estimation}
\label{subsec:EmpiricalLipschitz}
To evaluate the tightness of the RNN-SDP estimation \eqref{eq:FinalOptim} with \eqref{eq:Qdef} or \eqref{eq:QLocalSlope}, we compare the derived upper bounds to two empirical methods, that provide a lower bound of the Lipschitz constant: a naive \emph{random exploration} and a more targeted \emph{active exploration}.

\paragraph{Random Exploration}
A straightforward approach is to estimate the Lipschitz constant by evaluating the network on randomly generated input pairs, with the expectation that some may approximate the worst-case scenario. Since such cases typically occupy only a small subset of the input space, they are unlikely to be identified through naive sampling. Consequently, this method generally underestimates the true constant, with the degree of underestimation depending on the number and diversity of samples. Nevertheless, it provides a guaranteed lower bound on the Lipschitz constant and can serve as a useful consistency check.

We generate random input sequences uniformly from $[-1,1]$ and perturb them  
with additive Gaussian noise of variance $10^{-3}$ and mean zero.  
For each pair of original and perturbed inputs, we evaluate the empirical Lipschitz constant  
using a rearranged form of~\eqref{eq:LispschitzRNN}:
\begin{equation}
	L_{\mathrm{emp}} = \frac{\|\mathbf{y}_{N,2} - \mathbf{y}_{N,1}\|_2}
	{\left\lVert
		\begin{pmatrix} \gls{x}\\ \mathbf{h}_0 \end{pmatrix}_2 -
		\begin{pmatrix} \gls{x} \\ \mathbf{h}_0 \end{pmatrix}_1
		\right\rVert_2}.
	\label{eq:EmpiricalLipschitz}
\end{equation}
The final estimate, denoted $\gls{L_rand}$ in the results,  
is the maximum value observed across all evaluated pairs.  
In our experiments, we tested $10^7$ independent random samples.

\paragraph{Active Exploration}
Instead of relying on chance, we can actively search for inputs that maximize  
the Lipschitz constant. In this method, both the base input and its perturbation  
are treated as optimization variables. We perform gradient ascent  
on $L_{\mathrm{emp}}$ by minimizing $\mathcal{L} = -L_{\mathrm{emp}}$  
from~\eqref{eq:EmpiricalLipschitz}.  To reduce the likelihood of converging to a local maximum,  
we repeat the optimization from five different random initializations,  
retaining the best result as the empirical constant.  

For comparison with both bounded and unbounded RNN-SDP settings,  
we also implement a bounded variant: the inputs are passed through  
a scaled $\tanh(\cdot)$ to constrain them to $[-1,1]$,  
and perturbations are constrained to $[-10^{-3},10^{-3}]$.  
The smooth bounding ensures differentiability for stable convergence.  
Training stops if no improvement is seen for 10 consecutive epochs.  

The evaluation results are presented in the next section.

\section{Results}
\label{sec:Results}
In this section, we evaluate the tightness of the Lipschitz bounds obtained with RNN-SDP. Specifically, we compare the certified upper bounds to empirically derived lower bounds to assess estimation accuracy. The evaluation was conducted on 100 RNNs trained as described in Section~\ref{subsec:Training}.

\begin{figure}[!t]
	\begin{center}
\begin{tikzpicture}

\definecolor{darkgray176}{RGB}{176,176,176}
\definecolor{green01270}{RGB}{0,127,0}
\definecolor{lightgray204}{RGB}{204,204,204}

\begin{axis}[
width=7cm,
height=5cm,
scale only axis=true, 
trim axis left, trim axis right,
ylabel style={at={(axis description cs:0,0.5)}, anchor=south , yshift=2em},
xlabel style={at={(axis description cs:0.5,0)}, anchor=north, yshift=-1.5em},
scaled y ticks=false,                 
yticklabel style={
	/pgf/number format/fixed,
	/pgf/number format/precision=2,     %
	text width=2.5em, align=right       %
},	
legend cell align={center},
legend image post style={xscale=0.5}, 
legend style={
  fill opacity=0.8,
  draw opacity=1,
  text opacity=1,
  at={(0.5,1.02)},
  anchor=south,
  draw=lightgray204
},
legend columns=-1,
tick align=outside,
tick pos=left,
x grid style={darkgray176},
xlabel={Number of Unfolds [-]},
xmin=1, xmax=100,
xtick style={color=black},
y grid style={darkgray176},
ylabel={Estimated $L$ [-]},
ymin=0.0932393055083549, ymax=0.836498191220947,
ytick style={color=black}
]
\addplot [semithick, blue]
coordinates {%
(1,0.802713696415829)
(2,0.717743708791982)
(5,0.707633092514071)
(10,0.698960881207017)
(20,0.694866397108118)
(30,0.694537209958568)
(40,0.68427732717857)
(50,0.689285674529964)
(60,0.70483620286463)
(70,0.690643072122049)
(80,0.695636316986442)
(90,0.692780228447493)
(100,0.699508875636711)
};
\addlegendentry{$L_{RNN}$}
\addplot [semithick, blue,dashed]
coordinates {%
(1,0.80068313583425)
(2,0.712440258868916)
(5,0.700576148600867)
(10,0.690435171114109)
(20,0.686224746422341)
(30,0.68636979767142)
(40,0.675823206601541)
(50,0.681044902700954)
(60,0.695962979550564)
(70,0.682298759832128)
(80,0.686697941312006)
(90,0.683117815093633)
(100,0.690780675305647)
};
\addlegendentry{$L_{RNN,b}$}
\addplot [semithick, red]
coordinates {%
(1,0.795372080206871)
(2,0.657637029290199)
(5,0.587129133045673)
(10,0.552319157719612)
(20,0.546991426646709)
(30,0.546439421772957)
(40,0.540871004462242)
(50,0.545902815759182)
(60,0.550123463273048)
(70,0.547715112268925)
(80,0.543431069850922)
(90,0.5481524553895)
(100,0.552561281621456)
};
\addlegendentry{$L_{act}$}
\addplot [semithick, red,dashed]
coordinates {%
(1,0.789128710031509)
(2,0.647140007615089)
(5,0.574289436340332)
(10,0.53766642421484)
(20,0.533045840859413)
(30,0.53084503531456)
(40,0.526415832340717)
(50,0.531813459992409)
(60,0.534052007496357)
(70,0.532545029520988)
(80,0.530864969491959)
(90,0.530774666368961)
(100,0.536757406592369)
};
\addlegendentry{$L_{act,b}$}
\addplot [semithick, green01270]
coordinates {%
	(1,0.648987036347389)
	(2,0.46416196167469)
	(5,0.413607029020786)
	(10,0.35201996743679)
	(20,0.264836756139994)
	(30,0.222926297187805)
	(40,0.194851636439562)
	(50,0.17517625272274)
	(60,0.16006121173501)
	(70,0.153048092275858)
	(80,0.141147671565413)
	(90,0.131985377818346)
	(100,0.127023800313473)
};
\addlegendentry{$L_{Rand}$}
\end{axis}

\end{tikzpicture}
	\end{center}
	\caption{Upper bounds on the Lipschitz constant estimated by RNN-SDP (\gls{L_RNN},$\gls{L_RNN}_{,b}$) and the empirical lower bounds, derived by active exploration (\gls{L_act}, $\gls{L_act}_{,b}$) and random exploration (\gls{L_rand}). The subscript $b$ indicates the bounded version of an approach. The y-axis shows the estimated Lipschitz constant, based on the average of 100 NNs. The x-axis shows the number of times the RNN has been unrolled.}
	\label{fig:LipDevelopment}
\end{figure}
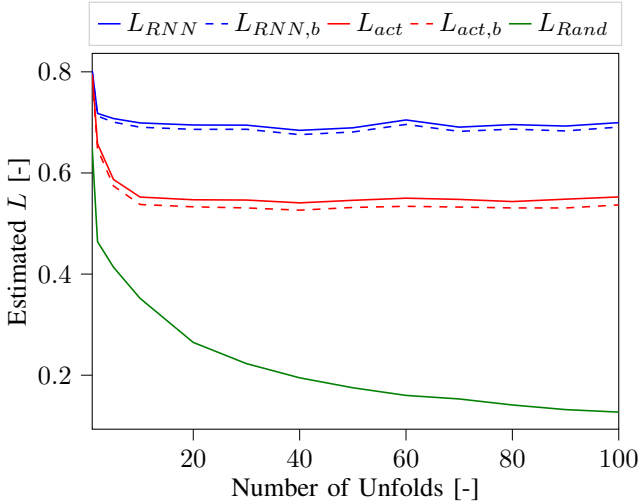

The estimated Lipschitz constants over sequence lengths from 1 to 100 are shown in Fig.~\ref{fig:LipDevelopment}. Five methods are compared: the RNN-SDP estimation ($L_{\mathrm{RNN}}$), its bounded-input variant ($L_{\mathrm{RNN,b}}$), active search ($L_{\mathrm{act}}$), and its bounded-input variant ($L_{\mathrm{act,b}}$), and random sampling ($L_{\mathrm{Rand}}$). Each curve reports the mean estimate across 100 trained RNNs.

\begin{figure}[!t]
	\begin{center}
\begin{tikzpicture}

\definecolor{darkgray176}{RGB}{176,176,176}
\definecolor{lightgray204}{RGB}{204,204,204}

\begin{axis}[
width=7cm,
height=5cm,
scale only axis=true, 
trim axis left, trim axis right,
ylabel style={at={(axis description cs:0,0.5)}, anchor=south , yshift=2em},
xlabel style={at={(axis description cs:0.5,0)}, anchor=north, yshift=-1.5em},
scaled y ticks=false,                 
yticklabel style={
	/pgf/number format/fixed,
	/pgf/number format/precision=2,     %
	text width=2.5em, align=right       %
},	
legend cell align={center},
legend style={
  fill opacity=0.8,
  draw opacity=1,
  text opacity=1,
  at={(0.5,1.02)},
  anchor=south,
  draw=lightgray204
},
legend columns=-1,
tick align=outside,
tick pos=left,
x grid style={darkgray176},
xlabel={Timestep [-]},
xmin=81, xmax=100,
xtick style={color=black},
y grid style={darkgray176},
ylabel={Input [-]},
ymin=-0.928851628303528, ymax=1.09180130958557,
ytick style={color=black}
]
\addplot [semithick, blue]
coordinates {%
(51,0.139288887381554)
(52,-0.154639884829521)
(53,-0.239211663603783)
(54,0.0527912080287933)
(55,-0.0434819869697094)
(56,0.193974882364273)
(57,-0.668634057044983)
(58,0.0605332367122173)
(59,-0.318234592676163)
(60,-0.726779937744141)
(61,-0.795718729496002)
(62,0.767147481441498)
(63,0.134263813495636)
(64,0.835118651390076)
(65,-0.693366169929504)
(66,0.455395996570587)
(67,0.636922657489777)
(68,-0.200200960040092)
(69,-0.44258326292038)
(70,-0.63238650560379)
(71,0.261184722185135)
(72,0.782129168510437)
(73,-0.63788366317749)
(74,-0.837003707885742)
(75,-0.47315126657486)
(76,-0.657785177230835)
(77,-0.835729718208313)
(78,-0.0181237980723381)
(79,0.394876450300217)
(80,0.20938116312027)
(81,0.0596797876060009)
(82,0.815659999847412)
(83,-0.807185351848602)
(84,0.477001905441284)
(85,0.664529919624329)
(86,-0.215422496199608)
(87,-0.00598288420587778)
(88,0.996565401554108)
(89,0.708532691001892)
(90,0.671814680099487)
(91,0.69593608379364)
(92,0.759679973125458)
(93,0.730879902839661)
(94,0.702684760093689)
(95,0.809206604957581)
(96,0.618013262748718)
(97,0.419373601675034)
(98,-0.0398333966732025)
(99,-0.292039602994919)
(100,-0.415608704090118)
};
\addlegendentry{input}
\addplot [semithick, red,dashed]
coordinates {%
(51,0.139288887381554)
(52,-0.154639884829521)
(53,-0.239211663603783)
(54,0.0527912080287933)
(55,-0.0434819869697094)
(56,0.193974882364273)
(57,-0.668634057044983)
(58,0.0604326389729977)
(59,-0.318234771490097)
(60,-0.726779937744141)
(61,-0.795718729496002)
(62,0.767147481441498)
(63,0.134263813495636)
(64,0.835118651390076)
(65,-0.693366169929504)
(66,0.455395996570587)
(67,0.636922657489777)
(68,-0.200200960040092)
(69,-0.44258326292038)
(70,-0.632467269897461)
(71,0.261184811592102)
(72,0.782128810882568)
(73,-0.63644802570343)
(74,-0.837003767490387)
(75,-0.473151206970215)
(76,-0.657785356044769)
(77,-0.835731387138367)
(78,-0.018130574375391)
(79,0.394864827394485)
(80,0.209363222122192)
(81,0.0596476495265961)
(82,0.815612852573395)
(83,-0.807212829589844)
(84,0.476919621229172)
(85,0.66464239358902)
(86,-0.215010687708855)
(87,-0.00572576327249408)
(88,0.99995344877243)
(89,0.714221239089966)
(90,0.681208431720734)
(91,0.710122764110565)
(92,0.780497014522552)
(93,0.760125637054443)
(94,0.741061568260193)
(95,0.858374714851379)
(96,0.676541924476624)
(97,0.481772720813751)
(98,0.00656857341527939)
(99,-0.262962967157364)
(100,-0.388382405042648)
};
\addlegendentry{disturbed input}
\end{axis}

\end{tikzpicture}
	\end{center}
	\caption{Normalized input trajectories of the first tank yielding the largest Lipschitz constant found by active exploration over a sequence of $100$ steps (only last 20 steps shown for clarity of presentation).
	}
	\label{fig:trajectory}
\end{figure}
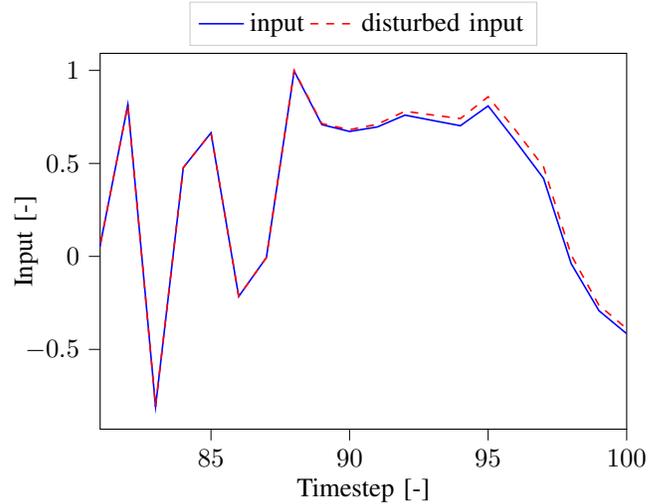

Across all methods, the estimates initially decrease with increasing sequence length and then converge to a plateau. This reflects the training objective, which enforced stability of the recurrent dynamics: as the RNN unrolls, the effect of initialization diminishes and sensitivity converges to a steady level. Convergence occurs around the 10\textsuperscript{th} unrolling for active-search methods and slightly earlier for the RNN-SDP. The effect is also evident in the worst-case trajectories identified by active exploration (Fig.~\ref{fig:trajectory}), where for a long sequence, inputs are initially identical and diverge only near the end of the horizon, where they can impact the final output.

The lower bound on the Lipschitz constant estimated via random sampling ($L_{\mathrm{Rand}}$) is consistently smaller than that obtained through active search, with the discrepancy increasing over longer horizons. This underestimation arises from the low probability of encountering worst-case trajectories through naive sampling, as the input space expands significantly with longer sequences. Consequently,  $L_{\mathrm{Rand}}$ represents a valid but overly loose lower bound that may \emph{misleadingly suggest low overall model sensitivity}, and is not suited for comparing the tightness of our method against, the active search method is used instead.

The proposed analytical method (RNN-SDP, $L_{\mathrm{RNN}}$) initially yields bounds that are very close to the active exploration results. At sequence length~1, the certified upper bound is less than 1\% larger than the empirical lower bound. With increasing sequence length, the bounds gradually loosen, stabilizing at roughly 30\% more conservative estimates. This degradation is likely due to compounding over-approximations from recursive bounding and local slope restrictions, rather than insufficient exploration in the active method. Nonetheless, the deviation remains moderate and consistent.

The high Lipschitz constant for short sequences also highlights the influence of the initialization, which is often not accounted for due to its effect diminishing after a short time \cite{revay2020convexparameterizationrobustrecurrent, Guo_10.1145/3583788.3583795}. However, for applications such as MPC, where the models are frequently reinitialized~\cite{Giuli10885933}, this effect must be accounted for in robustness analysis.

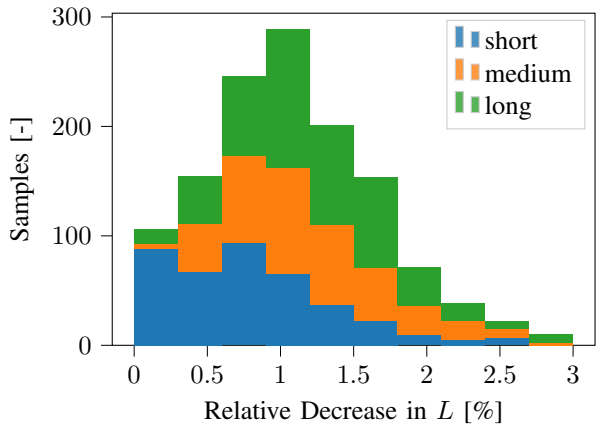
\begin{figure}[!t]
	\centering
\begin{tikzpicture}

\definecolor{darkgray176}{RGB}{176,176,176}
\definecolor{darkorange25512714}{RGB}{255,127,14}
\definecolor{forestgreen4416044}{RGB}{44,160,44}
\definecolor{lightgray204}{RGB}{204,204,204}
\definecolor{steelblue31119180}{RGB}{31,119,180}

\begin{axis}[
width=8cm,
height=6cm,
legend cell align={left},
legend style={fill opacity=0.8, draw opacity=1, text opacity=1, draw=lightgray204},
tick align=outside,
tick pos=left,
x grid style={darkgray176},
xlabel={Relative Decrease in $L$ [\%]},
xmin=-0.15, xmax=3.15,
xtick style={color=black},
y grid style={darkgray176},
ylabel={Samples [-]},
ymin=0, ymax=303.45,
ytick style={color=black}
]
\draw[draw=none,fill=steelblue31119180] (axis cs:2.77555756156289e-17,0) rectangle (axis cs:0.3,88);
\addlegendimage{ybar,ybar legend,draw=none,fill=steelblue31119180}
\addlegendentry{short}

\draw[draw=none,fill=steelblue31119180] (axis cs:0.3,0) rectangle (axis cs:0.6,67);
\draw[draw=none,fill=steelblue31119180] (axis cs:0.6,0) rectangle (axis cs:0.9,94);
\draw[draw=none,fill=steelblue31119180] (axis cs:0.9,0) rectangle (axis cs:1.2,66);
\draw[draw=none,fill=steelblue31119180] (axis cs:1.2,0) rectangle (axis cs:1.5,37);
\draw[draw=none,fill=steelblue31119180] (axis cs:1.5,0) rectangle (axis cs:1.8,22);
\draw[draw=none,fill=steelblue31119180] (axis cs:1.8,0) rectangle (axis cs:2.1,10);
\draw[draw=none,fill=steelblue31119180] (axis cs:2.1,0) rectangle (axis cs:2.4,5);
\draw[draw=none,fill=steelblue31119180] (axis cs:2.4,0) rectangle (axis cs:2.7,7);
\draw[draw=none,fill=steelblue31119180] (axis cs:2.7,0) rectangle (axis cs:3,0);
\draw[draw=none,fill=darkorange25512714] (axis cs:2.77555756156289e-17,88) rectangle (axis cs:0.3,93);
\addlegendimage{ybar,ybar legend,draw=none,fill=darkorange25512714}
\addlegendentry{medium}

\draw[draw=none,fill=darkorange25512714] (axis cs:0.3,67) rectangle (axis cs:0.6,111);
\draw[draw=none,fill=darkorange25512714] (axis cs:0.6,94) rectangle (axis cs:0.9,173);
\draw[draw=none,fill=darkorange25512714] (axis cs:0.9,66) rectangle (axis cs:1.2,162);
\draw[draw=none,fill=darkorange25512714] (axis cs:1.2,37) rectangle (axis cs:1.5,110);
\draw[draw=none,fill=darkorange25512714] (axis cs:1.5,22) rectangle (axis cs:1.8,71);
\draw[draw=none,fill=darkorange25512714] (axis cs:1.8,10) rectangle (axis cs:2.1,36);
\draw[draw=none,fill=darkorange25512714] (axis cs:2.1,5) rectangle (axis cs:2.4,22);
\draw[draw=none,fill=darkorange25512714] (axis cs:2.4,7) rectangle (axis cs:2.7,15);
\draw[draw=none,fill=darkorange25512714] (axis cs:2.7,0) rectangle (axis cs:3,2);
\draw[draw=none,fill=forestgreen4416044] (axis cs:2.77555756156289e-17,93) rectangle (axis cs:0.3,106);
\addlegendimage{ybar,ybar legend,draw=none,fill=forestgreen4416044}
\addlegendentry{long}

\draw[draw=none,fill=forestgreen4416044] (axis cs:0.3,111) rectangle (axis cs:0.6,154);
\draw[draw=none,fill=forestgreen4416044] (axis cs:0.6,173) rectangle (axis cs:0.9,246);
\draw[draw=none,fill=forestgreen4416044] (axis cs:0.9,162) rectangle (axis cs:1.2,289);
\draw[draw=none,fill=forestgreen4416044] (axis cs:1.2,110) rectangle (axis cs:1.5,201);
\draw[draw=none,fill=forestgreen4416044] (axis cs:1.5,71) rectangle (axis cs:1.8,154);
\draw[draw=none,fill=forestgreen4416044] (axis cs:1.8,36) rectangle (axis cs:2.1,71);
\draw[draw=none,fill=forestgreen4416044] (axis cs:2.1,22) rectangle (axis cs:2.4,38);
\draw[draw=none,fill=forestgreen4416044] (axis cs:2.4,15) rectangle (axis cs:2.7,22);
\draw[draw=none,fill=forestgreen4416044] (axis cs:2.7,2) rectangle (axis cs:3,10);
\end{axis}

\end{tikzpicture}
	\caption{Histogram of the relative reduction in estimated Lipschitz constant, when using the bounding method presented in  Section \ref{subsec:InputRestrictions}. The top of the stacked bars represent the distribution of the improvement within the data set. The colors within each bar indicate from which category of sequence length the samples came: short(1-10), medium(10-50), long(50-100). Showing that while an average improvement of 1.1\% was achieved, short sequences on average improved less (0.8\%) then medium (1.2\%) and long sequences (1.2\%)}
	\label{fig:LHist}
\end{figure}

Bounding the input ($L_{\mathrm{RNN,b}}$, $L_{\mathrm{act,b}}$) produces only marginally smaller constants. On average, bounding reduced the analytical estimates by 1.1\%, with short (1–10), medium (10–50), and long (50–100) sequences improving by 0.8\%, 1.2\%, and 1.3\%, respectively (Fig.~\ref{fig:LHist}). This suggests that the most sensitive inputs are already concentrated in the range $[-1,1]$, which is consistent with the steepest slope of the $\tanh(\cdot)$ activation near the origin. While bounding somewhat counteracts the compounding looseness of the method, the effect is too small to justify moving from a globally valid to a locally normalized bound.

Overall, the analytical method provides sound and certifiable bounds. While about 30\% more conservative than active search for long sequences, it guarantees validity, which is essential in safety-critical applications. Importantly, the certified bounds deviate by less than 1\% from the largest empirically observed constants, highlighting both their reliability and practical utility, especially for worst case estimations. The results also emphasize the risks of relying on random sampling, which systematically fails to capture critical worst-case behaviors.

\section{Conclusion}

In this paper, we introduced RNN-SDP, a novel method for computing certified upper bounds on the Lipschitz constant of RNNs over finite horizons. The approach explicitly accounts for both input disturbances and initialization uncertainty, providing a practical framework for robustness analysis in control-oriented and safety certification settings. 

Our evaluation shows that the proposed upper bounds deviate by less than 1\% from empirical lower bounds for short sequences and remain within about 30\% conservatism for long sequences. This accuracy is valuable for short-horizon tasks such as real-time feedback control and anomaly detection, and sufficiently robust for long-horizon tasks such as multi-step trajectory planning. Combined with formal validity guarantees, the method is well suited for safety-critical applications. In contrast, a naive empirical approach such as random sampling was found to consistently underestimated the Lipschitz constant, illustrating the risks of relying on naive statistical estimation-methods. 

We further examined whether bounds could be tightened by incorporating input-domain constraints through local slope restrictions. While this modification yielded slight improvements, 1.1\% on average, the gains were too small to justify abandoning globally valid guarantees and adding additional complexity to the approach. 

Overall, the proposed framework provides a step towards certifiable robustness analysis of RNNs, offering informative and guaranteed bounds that can serve as a foundation for robust model predictive control and related applications.
\bibliographystyle{IEEEtran}
\bibliography{References}

{\appendix[Point-wise vs. Sequence Outputs in RNN Lipschitz Analysis]
\label{ap:SeqVsPoint}
As noted in Section~\ref{subsec:RNNDefinition}, the output of an RNN can be defined either in a 
\emph{sequence-to-sequence} or a \emph{sequence-to-point} fashion. In the first case, the model maps
an input sequence $\mathbf{u} = \bigl(\!\begin{smallmatrix}\gls{x} \\ \mathbf{h}_0\end{smallmatrix}\!\bigr)$
to the stacked outputs
\begin{equation}
	f_{\mathrm{seq}}(\mathbf{u}) \!=\!\!
	\begin{pmatrix}\mathbf{y}_1 \\ \vdots \\ \mathbf{y}_N \end{pmatrix}
	\!\!=\!\!
	\begin{pmatrix}
		\gls{0} & \gls{W_out} & & \gls{0} \\
		& & \ddots & \\
		\gls{0} & \gls{0} & & \gls{W_out}
	\end{pmatrix} \!\mathbf{z} \!+\!
	\begin{pmatrix}\mathbf{b}_{\mathrm{out}}\! \\ \vdots \\ \mathbf{b}_{\mathrm{out}}\!\end{pmatrix},
\end{equation}
while in the sequence-to-point case only the final prediction is taken:
\begin{equation}
	f_{\mathrm{point}}(\mathbf{u}) = \mathbf{y}_N 
	= \begin{pmatrix}\gls{0} & W_{\mathrm{out}}\end{pmatrix}\mathbf{z} + \mathbf{b}_{\mathrm{out}}.
\end{equation}
Both definitions can in principle be used to estimate Lipschitz constants. However, for dynamical-system
applications, the sequence-based definition can underestimate sensitivity due to causality.
Specifically, the Lipschitz ratio for $f_{\mathrm{seq}}$ is
\begin{equation}
	L_{\mathrm{seq}} \!=\!
	\frac{\| f_{\mathrm{seq}}(\mathbf{u}_2) \shortminus f_{\mathrm{seq}}(\mathbf{u}_1)\|_2}
	{\|\mathbf{u}_2 \shortminus \mathbf{u}_1\|_2}
	\!=\!
	 \frac{\|(\Delta \mathbf{y}_1,\ldots,\Delta \mathbf{y}_N)\|_2}{\|(\Delta \mathbf{u}_1,\ldots,\Delta \mathbf{u}_N)\|_2}.
\end{equation}
For $N=2$, this becomes
\begin{equation}
	L_{\mathrm{seq}} = 
	\frac{\|(\Delta\mathbf{y}_1,\Delta\mathbf{y}_2)\|_2}{\|(\Delta\mathbf{u}_1,\Delta\mathbf{u}_2)\|_2}.
\end{equation}
Since $\mathbf{y}_1$ depends only on $u_1$, including $\Delta u_2$ in the denominator dilutes the
apparent sensitivity at step~1. For example, with
\[
(\Delta \mathbf{u}_1,\Delta \mathbf{u}_2)=(4,3), \qquad (\Delta \mathbf{y}_1,\Delta \mathbf{y}_2)=(5,0),
\]
we obtain
\begin{equation}
	L_{\mathrm{seq}} = \frac{\|(5,0)\|_2}{\|(4,3)\|_2} = 1.0, 
	\quad L_1 = \frac{5}{4}=1.25, \quad L_2=0.
\end{equation}
Here, the sequence-to-sequence definition undervalues the true stepwise sensitivity. In general, early
outputs can appear artificially less sensitive because the denominator contains future inputs that
cannot influence them.\\
To avoid this dilution, we adopt a pointwise analysis: for each horizon $t$, compute
\begin{equation}
	L_t = \sup_{\Delta u_{1:t}\neq 0}
	\frac{\|\mathbf{y}_t(u_{1:t}+\Delta \mathbf{u}_{1:t}) - \mathbf{y}_t(\mathbf{u}_{1:t})\|_2}{\|\Delta \mathbf{u}_{1:t}\|_2},
\end{equation}
and take $L = \max_{t=1,\ldots,N} L_t$. This definition respects causality and ensures that worst-case
step sensitivities are captured correctly.

}

\end{document}